\documentclass[aps,prd,preprintnumbers,amsmath,amssymb]{revtex4}

\usepackage{amsmath}
\usepackage{amssymb}
\usepackage{bm}
\usepackage{graphicx}
\usepackage{multirow}
\usepackage{subfigure}

\usepackage{textcomp}

\allowdisplaybreaks[1]

\begin{document}


\title{Vacuum polarization contribution to muon $g-2$ as an inverse problem}

\author{Hsiang-nan Li}
\affiliation{Institute of Physics, Academia Sinica,
Taipei, Taiwan 115, Republic of China}
\author{Hiroyuki Umeeda}
\affiliation{Institute of Physics, Academia Sinica,
Taipei, Taiwan 115, Republic of China}

\date{\today}

\begin{abstract}
We analyze the electromagnetic current correlator at an arbitrary photon invariant mass 
$q^2$ by exploiting its associated dispersion relation. The dispersion relation is turned 
into an inverse problem, via which the involved vacuum polarization 
function $\Pi(q^2)$ at low $q^2$ is solved with the perturbative input of $\Pi(q^2)$ 
at large $q^2$. It is found that the result for $\Pi(q^2)$, including its first derivative 
$\Pi^\prime(q^2=0)$, agrees with those from lattice QCD, and its imaginary part accommodates 
the $e^+e^-$ annihilation data. The corresponding hadronic vacuum polarization  
contribution $a^{\rm HVP}_\mu= (641^{+65}_{-63})\times 10^{-10}$ to the muon anomalous 
magnetic moment $g-2$, where the uncertainty arises from the variation of the perturbative 
input, also agrees with those obtained in other phenomenological and theoretical approaches.
We point out that our formalism is equivalent to imposing the analyticity constraint to
the phenomenological approach solely relying on experimental data, and can improve the precision
of the $a^{\rm HVP}_\mu$ determination in the Standard Model. 

\end{abstract}


%
%
%

\maketitle

%
%
%
\section{INTRODUCTION}

How to resolve the discrepancy between the theoretical prediction for the muon anomalous 
magnetic moment $a_\mu=(g_\mu-2)/2$ in the Standard Model and its experimental data 
has been a long standing mission. The major uncertainty in the former arises from 
the vacuum polarization function $\Pi(q^2)$ defined by an electromagnetic current correlator 
at a photon invariant mass $q^2$, to which various phenomenological and theoretical approaches
have been attempted. For instance, the measured cross section for $e^+e^-$ annihilation into 
hadrons has been employed to determine the hadronic vacuum polarization (HVP) contribution in 
a dispersive approach, giving $a^{\rm HVP}_\mu=(693.9\pm 4.0)\times 10^{-10}$ 
\cite{Davier:2019can} (see also $a^{\rm HVP}_\mu=(692.78 \pm 2.42)\times 10^{-10}$ in 
\cite{Keshavarzi:2019abf}). This value, consistent with earlier similar observations 
\cite{Davier:2010nc,Hagiwara:2011af,Davier:2017zfy,Keshavarzi:2019abf}, corresponds to a 
$3.3\sigma$ deviation between the Standard Model prediction for $a_\mu$ and the data \cite{PDG},
$a^{\rm exp}_\mu- a^{\rm SM}_\mu=(26.1\pm 7.9)\times 10^{-10}$. 
The above phenomenological determinations of $a^{\rm HVP}_\mu$, solely relying on experimental 
data, suffers a difficulty: the discrepancy among individual datasets, in particular between 
the BABAR and KLOE data in the dominant $\pi^+\pi^-$ channel, leads to additional systematic 
uncertainty \cite{Davier:2019can}. Therefore, theoretical estimates of the HVP contribution 
to the muon $g-2$ are indispensable, which have been performed mainly in 
lattice QCD (LQCD) (see \cite{Miura:2019xtd} for a recent review, and \cite{Borsanyi:2020mff} for a recent progress).
Results, such as $a^{\rm HVP}_\mu = (654\pm 32^{+21}_{-23})\times 10^{-10}$ in 
\cite{DellaMorte:2017dyu}, are comparable to those from the phenomenological approach.
It has been known that the finite volume in LQCD makes it unlikely to compute the vacuum polarization at 
low momenta with high statistics, for which a parametrization is always required to extrapolate 
lattice data.

In this paper we will calculate the vacuum polarization function in a novel method proposed recently
\cite{Li:2020xrz}, where a nonperturbative observable is extracted from its associated dispersion 
relation. Taking the $D$ meson mixing parameters as an example \cite{Li:2020xrz}, we separated
their dispersion relation for $D$ mesons of an arbitrary mass into a low mass piece and
a high mass piece, with the former being regarded as an unknown, and the latter 
being input from reliable perturbation theory. The evaluation
of the nonperturbative observable is then turned into an inverse problem: the observable at low mass
is solved as a "source distribution", which produces the "potential" at high mass. 
The resultant Fredholm integral equation allows the existence of multiple solutions
as a generic feature. However, it has been demonstrated that nontrivial solutions 
for the $D$ meson mixing parameters can be identified by specifying the physical 
charm quark mass, which match the data well. This
work implies that nonperturbative properties can be extracted from asymptotic QCD
by solving an inverse problem.

Here we will solve for the vacuum polarization function $\Pi(q^2)$ via an inverse problem, 
and derive the HVP contribution $a^{\rm HVP}_\mu$ to the muon $g-2$.
The electromagnetic current correlator is decomposed into three pieces according to the
quark composition of the $\rho$, $\omega$, and $\phi$ mesons. A dispersion relation is 
considered for each resonance, and converted into a Fredholm integral equation, which 
involves the unknown constant $\Pi(q^2=0)$ and the imaginary part ${\rm Im}\Pi(q^2)$
corresponding to the $e^+e^-\to (\rho,\omega,\phi)\to$ hadron spectra of 
nonperturbative origin. We solve the Fredholm equation with the perturbative input
of the leading order correlator at large $q^2$,
and select the solution, which best fits the $e^+e^-$ annihilation data for the resonance
spectra. The determined $\Pi(0)$, together with the 
resonance spectra at low $q^2$ and the perturbative input at high $q^2$, then yields 
$\Pi(q^2)$ from the dispersion relation. It will be shown that our predictions for 
$\Pi(q^2)$, including its first derivative $\Pi^\prime(q^2=0)$, and for $a^{\rm HVP}_\mu$ 
from the above three resonances
agree with those obtained in the literature.

We point out that simply inputting data into a dispersive approach does not automatically guarantee exact 
realization of the analyticity. When fitting the data, we search for the parameters involved in 
${\rm Im}\Pi(q^2)$ that satisfy the Fredholm equation, ie., the analyticity constraint, 
instead of tuning them arbitrarily. An intermediate impact of our 
formalism on other approaches is that one can impose the analyticity constraint to the 
conventional data-driven method.
That is, one may, for instance, check whether the dispersive integral of a dataset 
reproduces the perturbative $\Pi(q^2)$ at large $q^2$. It is then possible to discriminate 
the inconsistent datasets, such as the BABAR and KLOE data
mentioned above, so that the precision in the individual datasets can be fully exploited.
We will assess that such discrimination is achievable in principle, although the required precision 
for the perturbative input of $\Pi(q^2)$ goes beyond the scope of the present work.
%

The rest of the paper is organized as follows. In Sec.~II we present our formalism for 
extracting the nonperturbative vacuum polarization
function $\Pi(q^2)$ at low $q^2$, and solve the corresponding Fredholm equation.  
The similar procedure is extended to compute the slope
$\Pi'(0)$, that gives the leading contribution in the representation of
$\Pi(q^2)-\Pi(0)$ in terms of Pad\'e approximations \cite{MDB,CAT,MGK}, and serves as
a key ingredient in the "hybrid" approach proposed in \cite{Dominguez:2017yga}.
We evaluate the HVP contribution to the muon anomalous magnetic moment numerically in 
Sec.~III, and compare our prediction $a^{\rm HVP}_\mu= (641^{+65}_{-63})\times 10^{-10}$
from the $\rho$, $\omega$, and $\phi$ resonances,
where the uncertainty comes from the variation of the perturbative input,
with those from other phenomenological and LQCD approaches. Besides, we briefly demonstrate 
how to discriminate inconsistent datasets by imposing the analyticity constraint 
in light of attainable precise inputs in the future. Section IV is the conclusion.

\section{THE FORMALISM}

Start with the correlator 
\begin{eqnarray}
\Pi^{\mu\nu}_{\rm EM}(q)=i\int d^4x e^{iq\cdot x}\langle 0|
T[J^\mu(x) J^\nu(0)]|0\rangle=(q^\mu q^\nu-q^2g^{\mu\nu})\Pi_{\rm EM}(q^2),
\label{vac}
\end{eqnarray}
with the electromagnetic current $J^\mu(x)=\sum_f Q_f\bar q_f(x)\gamma^\mu q_f(x)$,
$Q_f$ being the charge of the quark $q_f$ with $f=u$, $d$, $s$. 
The leading order expression for the HVP contribution to the muon anomalous magnetic moment 
is written, in terms of the vacuum polarization function $\Pi_{\rm EM}(q^2)$, as 
\cite{Lautrup:1969uk,Lautrup:1971jf}
\begin{eqnarray}
a^{\rm HVP}_\mu&=&4\alpha_{\rm EM}^2
\int_0^1dx (1-x)\left[ \Pi_{\rm EM}(0)-\Pi_{\rm EM}\left(-\frac{x^2 m_\mu^2}{1-x}\right)\right],\label{amu4}
\end{eqnarray}
with the electromagnetic fine structure constant $\alpha_{\rm EM}$ and the muon mass $m_\mu$.
The first term can be set to $\Pi_{\rm EM}(0)=0$ \cite{deRafael:1993za} in the on-shell scheme 
for the QED renormalization, but is kept for generality, because it also receives 
nonperturbative QCD contribution. The behavior of $\Pi_{\rm EM}(-s)$ in the region with a large 
invariant mass squared $s$ has been known in perturbation theory. We will derive $\Pi_{\rm EM}(-s)$ 
in the low $s$ region, where the nonperturbative contributions from the $\rho$, 
$\omega$, and $\phi$ resonances dominate.

The vacuum polarization function obeys the dispersion relation
\begin{eqnarray}
-\frac{\Pi_{\rm EM}(-s)}{s}+\frac{\Pi_{\rm EM}(0)}{s}
&=&\frac{1}{\pi}\int_{\lambda}^\infty ds'\frac{{\rm Im}\Pi_{\mathrm{EM}}(s')}{s'(s'+s)},
\label{dis0}
\end{eqnarray}
$\lambda$ being a threshold. The function $\Pi_{\rm EM}(s)$ for large $s$ can be expressed as
\begin{eqnarray}
\Pi_{\rm EM}(s)=\displaystyle\sum_{f=u,d,s}Q_f^2\Pi(s, m_f),
\end{eqnarray}
$m_f$ being a light quark mass. The real parts of the functions $\Pi(s, m_f)$ at leading order 
are read off \cite{Kallen:1955fb} up to an overall normalization, 
\begin{eqnarray}
\Pi_{\rm OS}(-s, m_f)&=&\frac{5}{12\pi^2}-\frac{1}{\pi^2}\frac{m_f^2}{s}
-\frac{1}{2\pi^2}\sqrt{1+\frac{4m_f^2}{s}}\left(1-\frac{2m_f^2}{s}\right)
\tanh^{-1}\frac{1}{\sqrt{1+4m_f^2/s}},\nonumber\\
\Pi_{\rm \overline{ MS}}(-s, m_f)&=&\Pi_{\rm OS}(-s)-\frac{1}{4\pi^2}\ln\frac{\mu^2}{m_f^2},\nonumber\\
\Pi_{\rm {MS}}(-s, m_f)&=&\Pi_{\rm OS}(-s)-\frac{1}{4\pi^2}\left[\ln\frac{\mu^2}{m_f^2}+\ln(4\pi)-\gamma_E\right],
\label{Eq:Sc}
\end{eqnarray}
with $s>0$ in the on-shell, $\overline{\mathrm{MS}}$ and MS schemes for the QED renormalization, respectively. 
The imaginary part is given by \cite{Kallen:1955fb}
\begin{eqnarray}
\mathrm{Im}\Pi(s, m_f)&=&\begin{cases}
0, \quad s<4m_f^2\\
\frac{1}{4\pi}\sqrt{1-4m_f^2/s}(1+2m^2_f/s),\quad s\geq 4m_f^2.
\end{cases}\label{Eq:Im}
\end{eqnarray}
It is seen that the real parts $\Pi_{\rm EM}(-s)$ in the above schemes differ by the $s$-independent 
terms, which can be always absorbed into the redefinition of the unknown constant $\Pi_{\rm EM}(0)$ in 
Eq.~(\ref{dis0}). It is also clear that our result for $a^{\rm HVP}_\mu$ will not depend on the choice 
of a specific renormalization scheme, because the scheme dependence cancels between the two terms 
in Eq.~(\ref{amu4}). Hence, we will stick to the on-shell scheme, and omit the subscript OS
in the formulation below.

We decompose Eq.~(\ref{dis0}) into three separate dispersion 
relations labelled by $r=\rho$, $\omega$, $\phi$, and rewrite them as
\begin{eqnarray}
& &\int_{\lambda_r}^{\Lambda_r} ds'\frac{{\rm Im}\Pi_r(s')}{s'(s'+s)}-\pi\frac{\Pi_r(0)}{s}
=\Omega_r(s),\label{disx}\\
& &\Omega_r(s)\equiv -\pi \frac{\Pi_r(-s)}{s}-\int_{\Lambda_r}^{\infty} 
ds'\frac{{\rm Im}\Pi_r(s')}{s'(s'+s)},\label{disy}
\end{eqnarray}
where the thresholds are set to $\lambda_\rho=4m_{\pi^+}^2$, $\lambda_\omega=(2m_{\pi^+}+m_{\pi^0})^2$, 
and $\lambda_\phi=4m_{K^+}^2$, with the pion (kaon) mass $m_\pi$ ($m_K$). The separation scale 
$\Lambda_r$ will be determined later, which is expected to be large enough to justify the perturbative 
calculation of the imaginary part ${\rm Im}\Pi_r(s)$ in Eq.~(\ref{disy}).
Equation~(\ref{disx}) is then treated as an inverse problem, {\it i.e.}, a Fredholm integral equation,
where $\Omega_r(s)$ defined by Eq.~(\ref{disy}) for $s>\Lambda_r$ is an input, and ${\rm Im}\Pi_r(s)$ 
in the range $s<\Lambda_r$ is solved with 
the continuity of ${\rm Im}\Pi_r(s)$ at $s=\Lambda_r$. That is, the "source distribution" 
${\rm Im}\Pi_r(s)$ will be inferred from the "potential" $\Omega_r(s)$ observed outside 
the distribution. Equation~(\ref{disx}) can be regarded as a realization of the 
global quark-hadron duality postulated in QCD sum rules \cite{Shifman:2000jv}.

\begin{figure}
\includegraphics[scale=0.47]{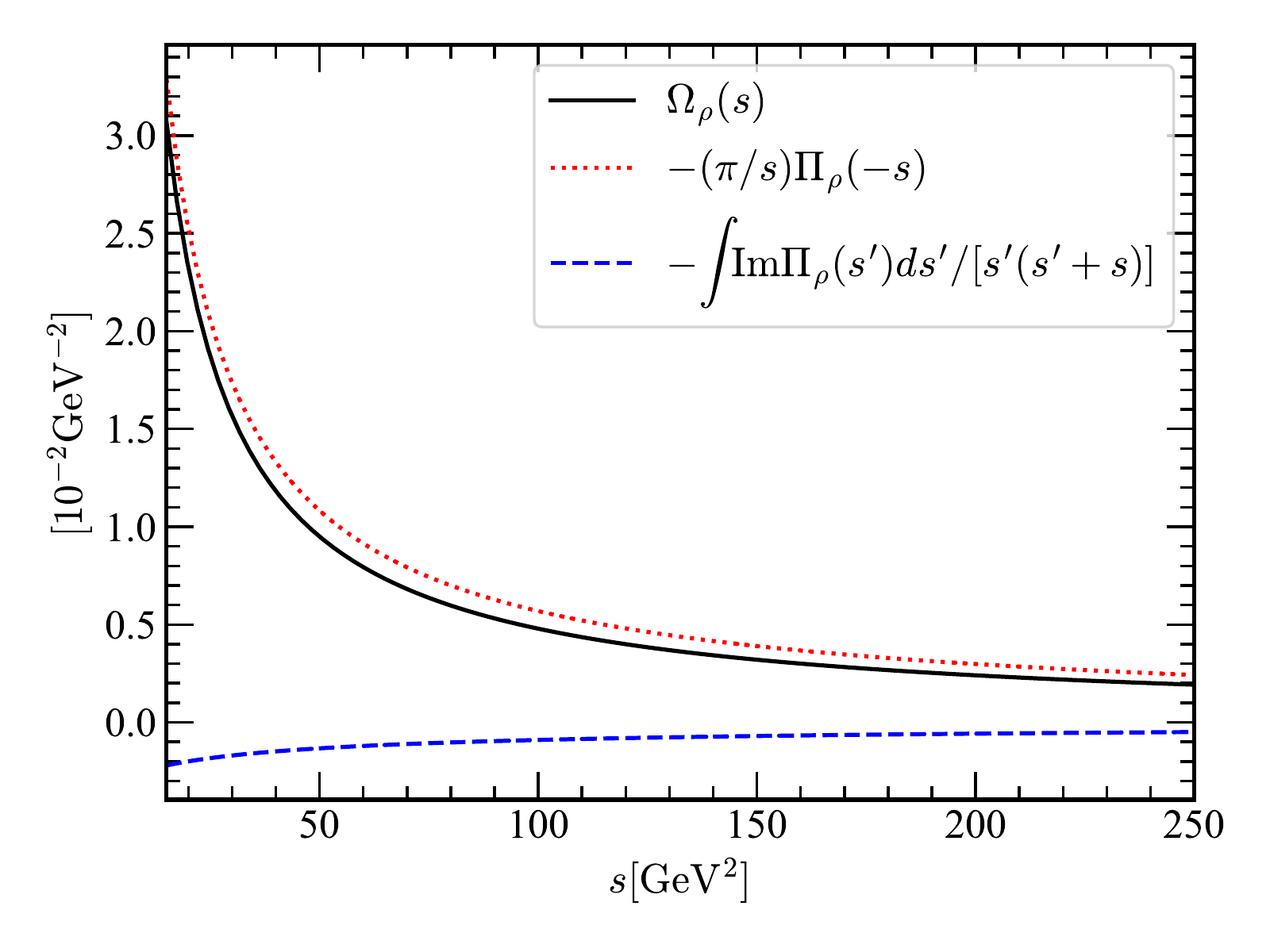}
\caption{\label{fig1}
$s$ dependence of $\Omega_\rho(s)$ with the input parameters $\Lambda_\rho=11.6\;\mathrm{GeV}^2$,
$m_u=2.16\;\mathrm{MeV}$, and $m_d=4.67\;\mathrm{MeV}$ in the on-shell scheme.}
\end{figure}

Both the real and imaginary parts of the input functions $\Pi_r(s)$ in $\Omega_r(s)$ 
are related to $\Pi(s, m_f)$ via 
\begin{eqnarray}
\Pi_\rho(s)=C_\rho \Pi(s, (m_u+m_d)/2),\;\;\;\;
\Pi_\omega(s)=C_\omega \Pi(s, (m_u+m_d)/2),\;\;\;\;
\Pi_\phi(s)=C_\phi \Pi(s, m_s),
\end{eqnarray}
with the charge factors $C_\rho=[(Q_u-Q_d)/\sqrt{2}]^2=1/2$, $C_\omega=[(Q_u+Q_d)/\sqrt{2}]^2=1/18$, and 
$C_\phi=Q_s^2=1/9$. The behaviors of $-\pi \Pi_\rho(-s)/s$, $-\int ds'{\rm Im}\Pi_\rho(s')/[s'(s'+s)]$, 
and $\Omega_\rho(s)$ in Eq.~(\ref{disy}) for the running masses $m_u=2.16$ MeV and 
$m_d=4.67$ MeV at the scale 2 GeV, and the separation scale $\Lambda_\rho=11.6$ GeV$^2$ are displayed in Fig.~\ref{fig1}. The behaviors
of the quantities for the $\omega$ and $\phi$ resonances, obtained with the replacements
of the quark masses ($m_s=93$ MeV), are similar.
Note that an inverse problem is usually ill-posed, and the ordinary discretization method to solve 
a Fredholm integral equation does not work. The discretized version of Eq.~(\ref{disx}) is 
in the form $\sum_i A_{ij}{\rm Im}\Pi_{j}-\pi\Pi_r(0)/s_i=\Omega_i$ with $A_{ij}\propto 1/[j(i+j)]$. 
It is easy to find that any two adjacent rows of the matrix $A$ approach to each other as the grid 
becomes infinitely fine. Namely, $A$ tends to be singular, and has no inverse. We stress 
that this singularity, implying no unique solution, should be appreciated actually. If
$A$ is not singular, the solution to Eq.~(\ref{disx}) will be unique, which must be
the perturbative results in Eqs.~(\ref{Eq:Sc}) and (\ref{Eq:Im}). It is the existence of multiple
solutions that allows possibility to account for the nonperturbative ${\rm Im}\Pi_r(s)$ 
in the resonance region.
After solving for $\Pi_r(0)$ together with ${\rm Im}\Pi_r(s)$ in the whole range of $s$, we derive 
$\Pi_r(-s)$ from the three dispersion relations, and $\Pi_{\rm EM}(-s)$ from their sum 
to be inserted into Eq.~(\ref{amu4}).

Knowing the difficulty to solve an inverse problem and the qualitative behavior of a resonance spectrum, we propose the parametrizations 
\begin{eqnarray}
{\rm Im}\Pi_\rho(s)&=&
\left\{\left(1-\frac{\lambda_\rho}{s}\right)
\frac{b^\rho_0
|1+\kappa s/(m_\omega^2-s-id_\omega)|^2}{(s-m_\rho^2)^2[1+z_2(s-m_\rho^2)^2]
+d_\rho^2[1+z_1(s-m_\rho^2)]}+c^\rho_0(s-\lambda_\rho)\right\}\theta(1-\lambda_\rho/s),\nonumber\\
{\rm Im}\Pi_\omega(s)&=&
\left[\left(1-\frac{\lambda_\omega}{s}\right)\frac{b^\omega_0}
{(s-m_\omega^2)^2+d_\omega^2}+c_0^\omega (s-\lambda_\omega)\right]
\theta(1-\lambda_\omega/s),\nonumber\\
{\rm Im}\Pi_\phi(s)&=&
\left[\left(1-\frac{\lambda_\phi}{s}\right)\frac{b^\phi_0}{(s-m_\phi^2)^2+d_\phi^2}+c_0^\phi (s-\lambda_\phi)\right]\theta(1-\lambda_\phi/s),\label{para}
\end{eqnarray}
according to \cite{Eidelman:1995ny,Lichard:2006ky}, where $d_r=m_r\Gamma_r$ is the
product of the meson mass $m_r$ and the width $\Gamma_r$. 
The parameter $b^r_0$ ($c^r_0$) describes the strength of the resonant (nonresonant) contribution,
and $\kappa$ characterizes the $\rho$-$\omega$ mixing effect.
We have adopted the same threshold for the $K^+K^-$, $K_SK_L$ and $\pi^+\pi^-\pi^0$ final states 
of $\phi$ decays for simplicity. 
For the denominator of the $\rho$ resonance in Eq.~(\ref{para}), we introduce the linear and quartic terms 
in $s-m_\rho^2$, which are motived by the Gounaris-Sakurai model \cite{GS}. We have verified that the 
gross shape of the Gounaris-Sakurai model for the resonance is reproduced with this simpler parametrization
in order to facilitate the numerical analysis below.
The parameters $z_1$ and $z_2$ lead to the effective width and mass of a $\rho$ meson.
This can be understood by completing the square of the denominator of the resonance term, with the quartic 
term being left aside first. The $z_1$ term then shifts the $\rho$ meson mass and width into 
$m_\rho^{\prime 2} = m_\rho^2-z_1d_\rho^2/2$ and $d_\rho^{\prime 2} = d_\rho^2(1-z_1^2d_\rho^2/4)$.
The approximation $z_2(s-m_\rho^{2})^4\approx z_2(s-m_\rho^{\prime 2})^4$ valid for $|s|\gg m_\rho^2$ 
will be assumed. We have confirmed that the quartic term is much 
less important than the quadratic term in the denominator even for $s\sim m_\rho^2$ and 
$z_1$ and $z_2$ determined later, so the approximation indeed holds well.

We have examined that the variations of the meson masses $m_r$ and widths $\Gamma_r$ and
the $\rho$-$\omega$ mixing parameter $\kappa$ change our results at $0.1\%$ level, so
$m_r$ and $\Gamma_r$ are set to their values in \cite{PDG}, and the mixing parameter is set 
to $\kappa=2.16\times 10^{-3}$ \cite{Davier:2019can}.
The free parameters $z_1$, $z_2$, $b^r_0$, $c^r_0$, $\Lambda_r$ and ${\rm Im}\Pi_r(0)$ 
are then tuned to best fit the input $\Omega_r(s)$ under
the continuity requirement from ${\rm Im}\Pi_r(s=\Lambda_r)$.
The separation scale $\Lambda_r$ introduces an end-point
singularity into $\Omega_r(s)$ in Eq.~(\ref{disy})
as $s'\to\Lambda_r$. To reduce the effect caused by this artificial singularity,
we consider $\Omega_r(s)$ from the range 15 GeV$^2<s<$ 250 GeV$^2$, in which
200 points $s_i$ are selected. 
We then search for the set of parameters, that minimizes the residual sum of square (RSS) 
\begin{eqnarray}
\sum_{i=1}^{200} \left|\int_{\lambda_r}^{\Lambda_r} ds'
\frac{{\rm Im}\Pi_r(s')}{s'(s'+s_i)}-\frac{\pi\Pi_r(0)}{s_i}
-\Omega_r(s_i)\right|^2.\label{dev}
\end{eqnarray}
Such a set of parameters corresponds to a solution of the Fredholm equation in Eq.~(\ref{disx})
in terms of the parametrizations in Eq.~(\ref{para}), namely, respects the analyticity constraint
most.

\section{NUMERICAL ANALYSIS}
\subsection{HVP Contribution}
\begin{figure}
\subfigure[]{\includegraphics[scale=0.47]{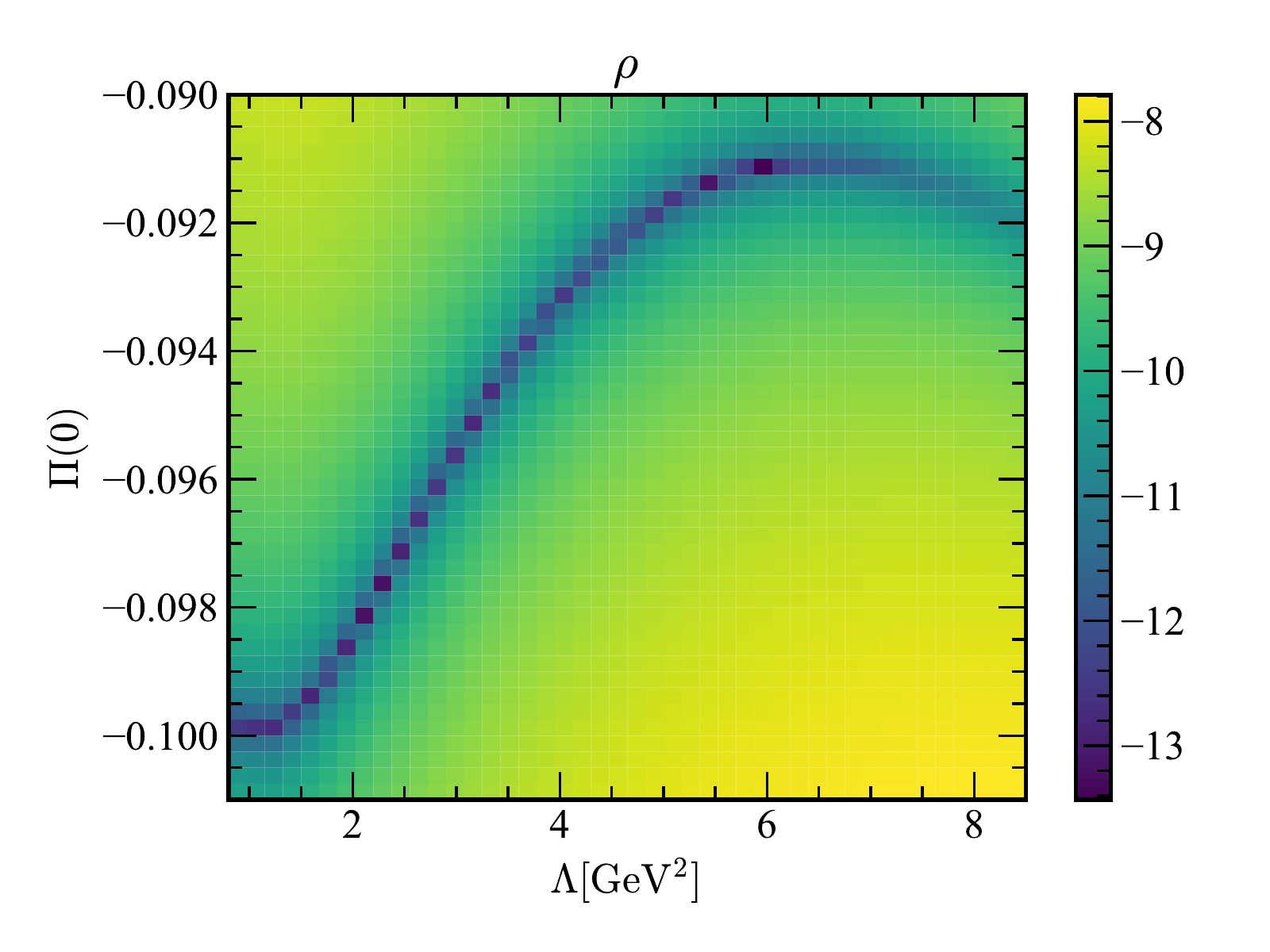}}
\subfigure[]{\includegraphics[scale=0.47]{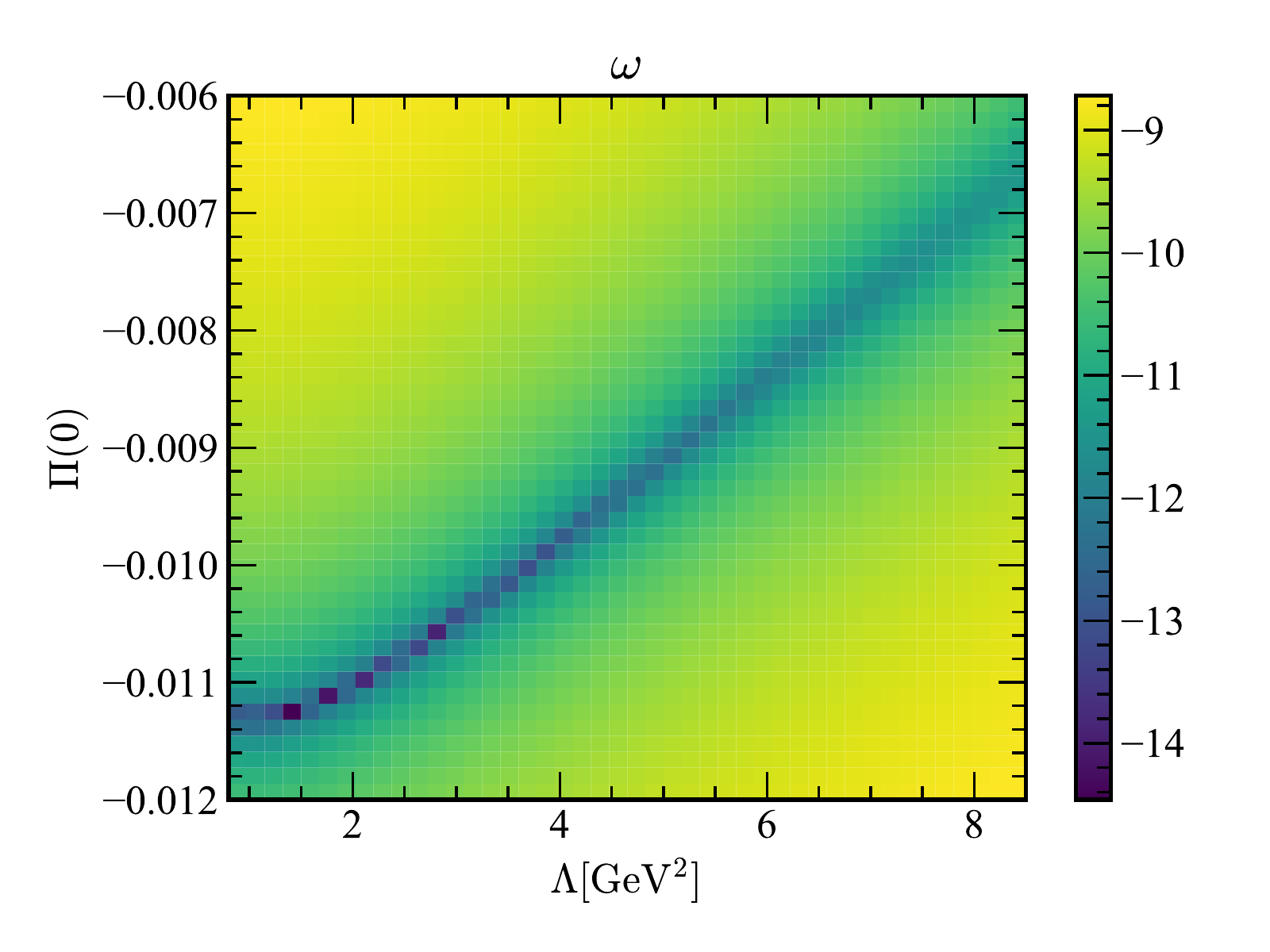}}
\subfigure[]{\includegraphics[scale=0.47]{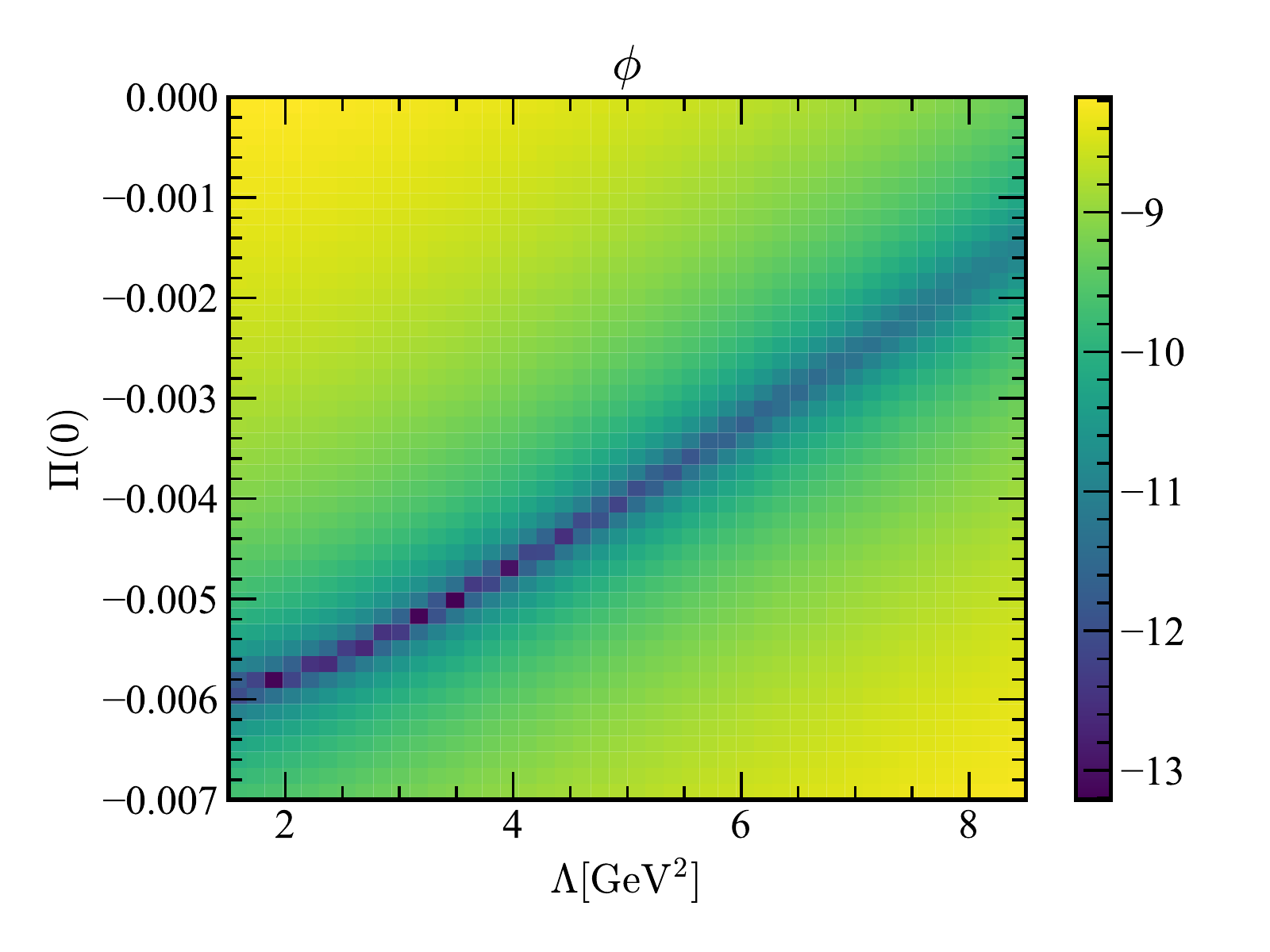}}
\caption{\label{fig2}
RSS minimum structures from the Fredholm equations for the (a) $\rho$, (b) $\omega$, and (c) $\phi$
resonances. The parameters $z_1=2.7$ $\mathrm{GeV}^{-2}$ and
$z_2= 0.532$ $\mathrm{GeV}^{-4}$ have been fixed for (a).
}
\end{figure}

The scanning over all the free parameters reveals the minimum distributions of the RSS 
defined in Eq.~(\ref{dev}), and typical distributions on the $\Lambda_r$-$\Pi_r(0)$ plane 
are displayed in Fig.~\ref{fig2}. The minima along the curve, having 
RSS about $10^{-12}$-$10^{-13}$ relative to $10^{-8}$ from outside the curve, hint the 
existence of multiple solutions. A value of $\Lambda_r$ represents the scale, at which the 
nonperturbative resonance solution starts to deviate from the perturbative input. This 
explains the dependence on $\Lambda_r$ of a solution. It is observed that the solutions for
$\Pi_r(0)$, including the sign and magnitude, fall in the same ballpark as LQCD results 
\cite{DellaMorte:2017dyu}. We then search for a solution along the RSS minimum distribution, 
which best accommodates the $e^+e^-$-annihilation data.
For the $\rho$ resonance spectrum, we consider the SND data for the process 
$e^+ e^- \to \pi^+ \pi^-$ from VEPP-2M collider 
in \cite{Achasov:2005rg}, which are consistent with those from all other collaborations 
as indicated by Fig.~5 in \cite{Davier:2019can}. 
It means that we are 
making a conservative prediction for the HVP contribution to the muon anomalous magnetic moment. 
We are guided by the data for the process $e^+ e^- \to \pi^+ \pi^- \pi^0$
through the $\omega$ resonance in \cite{Achasov:2003ir}. For the $\phi$ resonance, 
the SND data \cite{Achasov:2000am} are also adopted, which include the $e^+ e^- \to K^+ K^-$, $K_SK_L$, 
and $\pi^+ \pi^- \pi^0$ channels. We explain the fitting procedure for the $\rho$ resonance spectrum 
in more detail: because of the additional parameters $z_1$ and $z_2$ involved in this case, we first 
select a set of $z_1$ and $z_2$ values, perform the above fitting procedure to find the best
fit to the data, and then vary $z_1$ and $z_2$ to further improve the best fit.
The parameters $z_1=2.7$ $\mathrm{GeV}^{-2}$ and
$z_2= 0.532$ $\mathrm{GeV}^{-4}$ are obtained in this way, based on which 
Fig.~\ref{fig2}(a) is generated.

\begin{figure}
\includegraphics[scale=0.29]{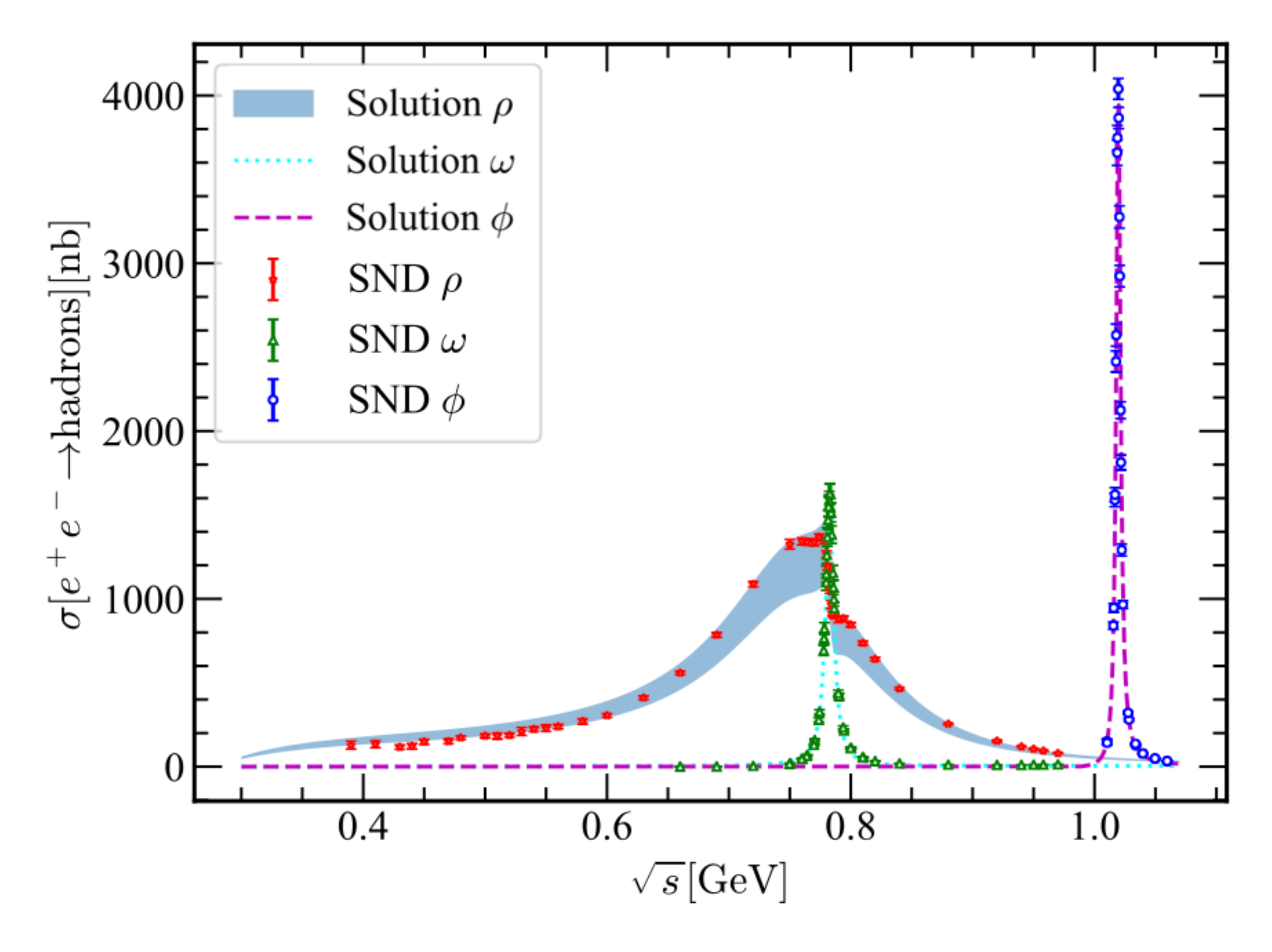}
\caption{\label{fig3}
Cross sections for $e^+e^-\to (\rho, \omega, \phi)\to\mathrm{hadrons}$ obtained as solutions
of the inverse problem. The SND data from VEPP-2M collider \cite{Achasov:2005rg, Achasov:2003ir, Achasov:2000am} are 
also exhibited for comparison. 
The data for the three modes $e^+e^-\to \phi\to \pi^+\pi^-\pi^0, K_SK_L$ and $K^+K^-$ have been combined with their uncertainties being added in quadrature.}
\end{figure}

Searching for the parameters along the RSS minimum distributions in Fig.~\ref{fig2}, 
we find that the parameters
\begin{eqnarray}
\rho&:&\; \Lambda_{\rho}= 11.6\;\mathrm{GeV}^2,\quad
b_0^\rho= 2.97\times 10^{-3}\;\mathrm{GeV}^4,\quad
c_0^{\rho}= 3.45\times 10^{-3}\;\mathrm{GeV}^{-2},\quad
\Pi_\rho(0)= -0.0954,\nonumber\\
\omega&:&\;\Lambda_{\omega}= 2.8\;\mathrm{GeV}^2,\quad
b_0^\omega=1.72\times 10^{-5}\;\mathrm{GeV}^4,\quad
c_0^{\omega}= 1.51\times 10^{-3}\;\mathrm{GeV}^{-2},\quad
\Pi_\omega(0)=-0.00953,\nonumber\\
\phi&:&\;\Lambda_{\phi}= 3.2\;\mathrm{GeV}^2,\quad
b_0^\phi= 3.90\times 10^{-4}\;\mathrm{GeV}^4 ,\quad
c_0^{\phi}= 3.95\times 10^{-3}\;\mathrm{GeV}^{-2},\quad
\Pi_\phi(0)=-0.00520,
\label{Eq:fitres}
\end{eqnarray}
best fit the $e^+e^-$-annihilation data
through the $\rho, \omega$ and $\phi$ resonances. The values of $\Lambda_r$ in Eq.~(\ref{Eq:fitres})
are large enough for justifying the perturbative evaluation of the input $\Omega_r(s)$.
Note that the above parameters follow the
correlation demanded by the perturbative input via the Fredholm equation, and are not 
completely free. This correlation, originating from the analyticity of the vacuum polarization, 
distinguishes our approach from the phenomenological one
\cite{Davier:2010nc,Hagiwara:2011af,Davier:2017zfy,Davier:2019can,Keshavarzi:2019abf},
in which the free parameters are solely determined by data fitting. 
We emphasize that a sensible resonance spectrum should be a solution of the Fredholm
equation, {\it i.e.}, respect the analyticity of the vacuum polarization. Therefore,
one may check whether a dataset obeys the Fredholm equation, {\it i.e.}, whether its dispersive 
integral reproduces the perturbative vacuum polarization function at large $s$, before it is 
employed in the phenomenological approach. This check will help discriminating inconsistent 
datasets, such as the BABAR and KLOE data mentioned before, and enhancing the precision of the
obtained hadronic contribution to the muon $g-2$.

The predicted cross sections corresponding to the sets of parameters in Eq.~(\ref{Eq:fitres})
are shown in Fig.~\ref{fig3}, which agree with the measured $\omega$ and
$\phi$ resonance spectra well, but deviate from the $\rho$ spectrum slightly. The agreement is
nontrivial, viewing the correlation imposed by the analyticity constraint on the parameters.
A parametrization more sophisticated than Eq.~(\ref{para}), {\it e.g.}, the one proposed in 
\cite{Davier:2019can} below the threshold of the inelastic scattering may improve
the agreement in the $\rho$ channel. However, we will not attempt an exact fit, since the SND data are 
just one of the many available datasets, and subject to the scrutinization of the analyticity
constraint to be elaborated in Sec.~III C. Instead, we investigate 
whether the theoretical uncertainty in the present analysis can explain the deviation. 
Higher order QCD corrections to the perturbative input cause about 
$\alpha_s/\pi\sim 10\%$ variation at the scale of $\Lambda_r$ around few GeV$^2$ \cite{Chetyrkin:1996cf}. 
As a test, we increase and decrease the perturbative input in Eq.~(\ref{disy}) by 10\%,
and estimate the errors associated with this variation by repeating the above procedure
for the same fixed values of $z_1$ and $z_2$.
We pick up the minima of RSS corresponding to $\Omega_i$ with $+10\%$ and $-10\%$ variations, 
$i=\rho, \omega$ and $\phi$, at $\Lambda_i$ in Eq.~(\ref{Eq:fitres}). 
The parameters $b_0^i, c_0^i$ and $\Pi_i(0)$ read off from the above minima then lead to 
the error bands in Fig.~\ref{fig3}, although the bands associated with the $\omega$ and $\phi$ spectra 
are too thin to be seen. It is found that most data for the $\rho$ spectrum are covered 
(the recent SND data for the $\rho$ spectrum \cite{Achasov:2020iys} 
are also covered, while the new result is in conflict with both the BABAR and KLOE experiments), 
except the tail part at low $s$, which gives a minor contribution to $a^{\rm HVP}_\mu$. 
It implies that the estimate of the theoretical uncertainty through the variation of the perturbative
input is relevant. 
Certainly, different choices of the parametrizations for the resonance spectra 
may also cause theoretical uncertainty. Because our results have matched the data satisfactorily, 
we do not take into account this source of uncertainty here.

\begin{figure}
\includegraphics[scale=0.6]{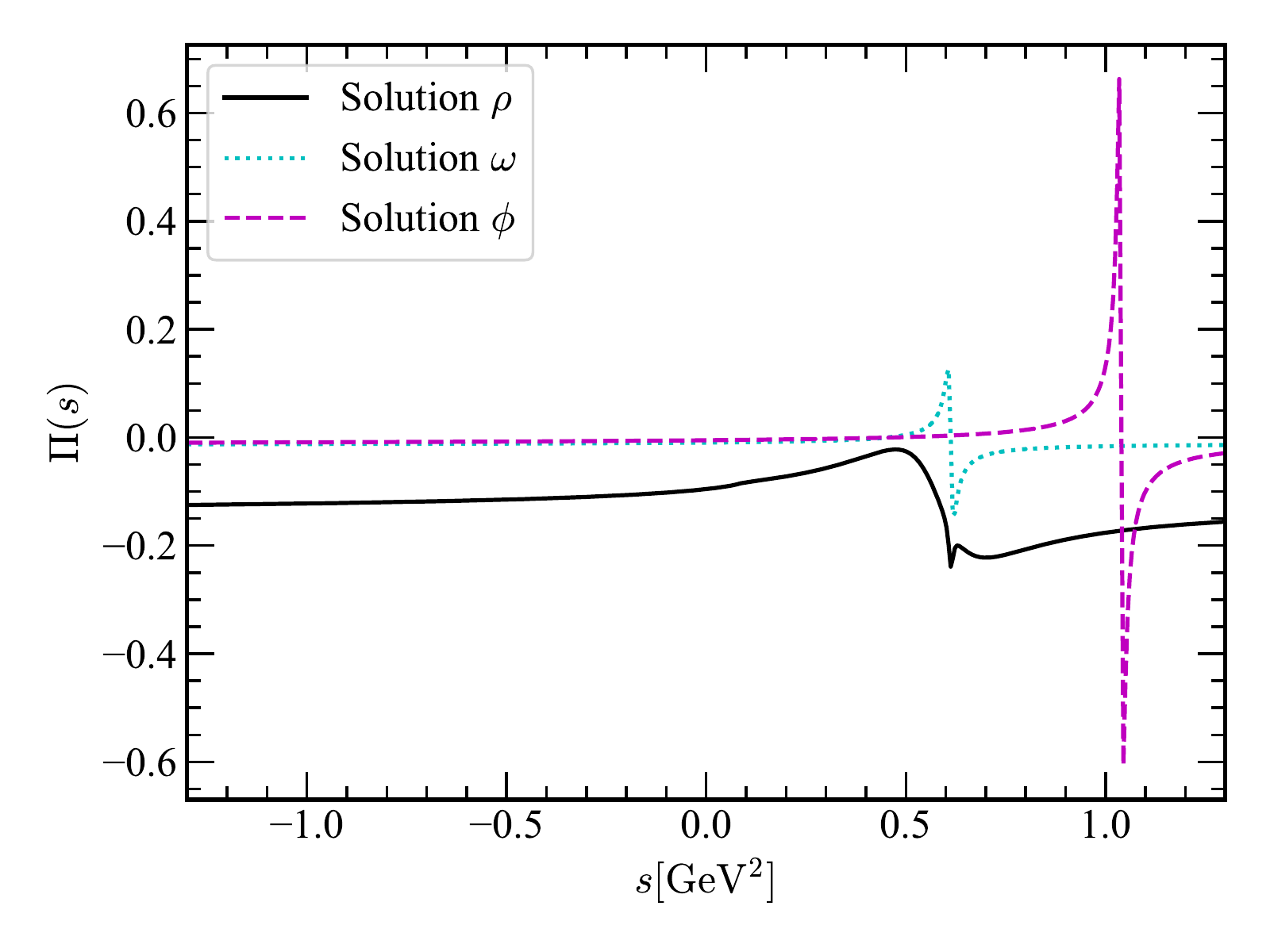}
\caption{\label{fig4}
Vacuum polarization functions associated with the $\rho, \omega$ and $\phi$ resonances obtained 
as solutions of the inverse problem.}
\end{figure}

\begin{figure}
\includegraphics[scale=0.28]{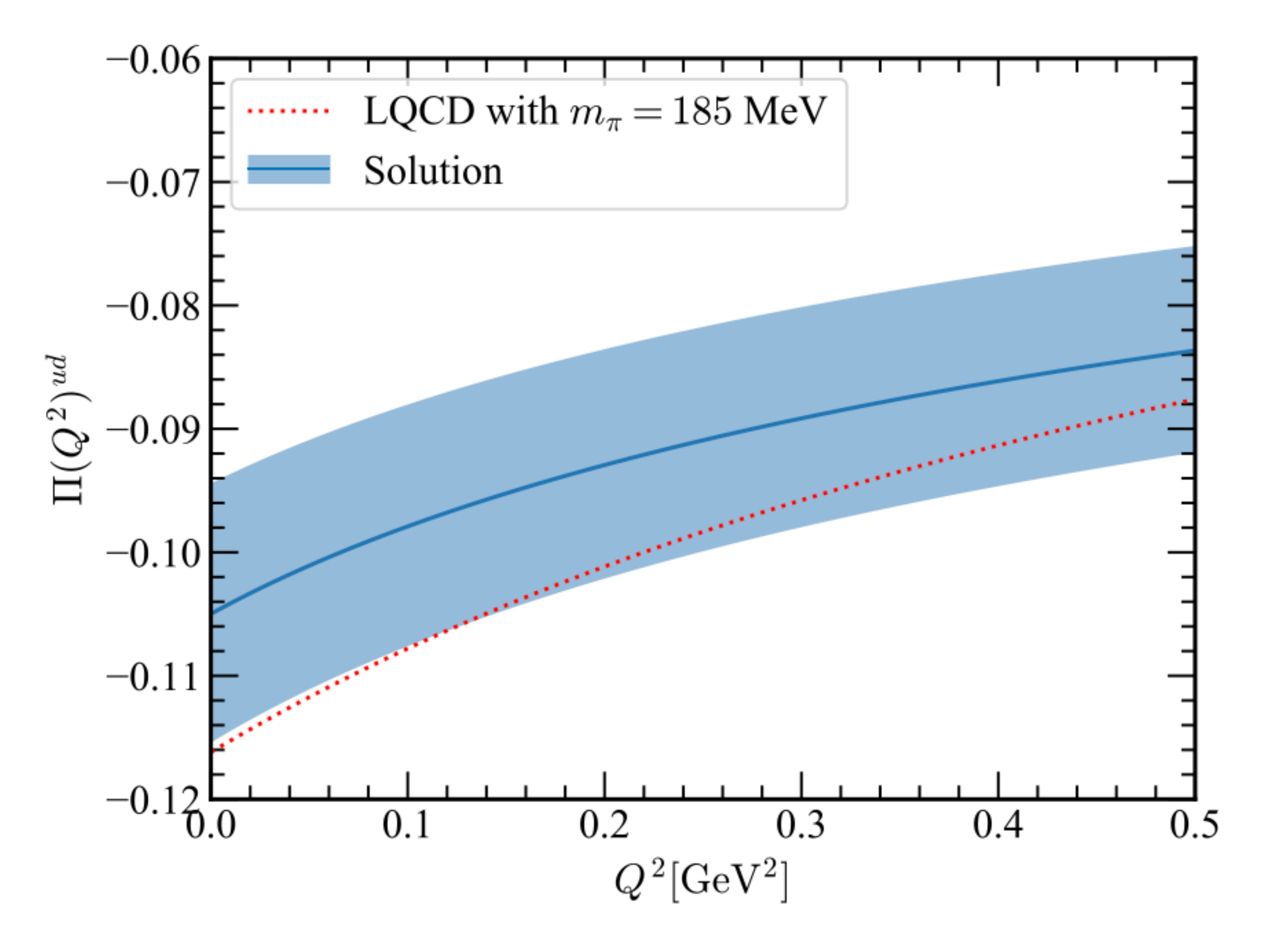}
\caption{\label{fig5}
Comparison of the predicted $\Pi(Q^2)^{ud}$ in the Euclidean momentum space with the LQCD result 
\cite{DellaMorte:2017dyu}. See text for the definition of $\Pi(Q^2)^{ud}$.
}
\end{figure}

Once the imaginary part ${\rm Im}\Pi_r(s)$ at low $s$ is derived, its behavior in the whole
$s$ range is known (with the perturbative input at high $s$), and 
the real part $\Pi_r(-s)$ can be calculated from Eq.~(\ref{dis0}).
The behaviors of the vacuum polarization functions in both the space-like $s<0$ and 
time-like $s>0$ regions are presented in Fig.~\ref{fig4}. The oscillations of the curves
ought to appear, when the photon invariant mass crosses physical resonance masses. 
The predicted vacuum polarization function from the $u$ and $d$ quark currents, {\it i.e.}, the $\rho$ and 
$\omega$ meson contributions is exhibited in Fig.~\ref{fig5}. In order to compare our 
result with $\Pi(Q^2)^{ud}$ in LQCD \cite{DellaMorte:2017dyu}, where a photon invariant mass 
is defined in the Euclidean momentum space, we have converted Eq.~(\ref{dis0}) into
$\Pi(Q^2)^{ud}=\Pi_{\rho}(0)+\Pi_{\omega}(0)
+(Q^2/\pi)\int ds^\prime[\Pi_{\rho}(s^\prime)+\Pi_{\omega}(s^\prime)]/[s^\prime(s^\prime+Q^2)]$.
It is obvious that our prediction for $\Pi(Q^2)^{ud}$ agrees with the LQCD one corresponding to the 
pion mass $m_\pi=185$ MeV within the 10\% theoretical uncertainty. The LQCD results show the tendency 
of decreasing with the pion mass, so a better agreement is expected, if a further lower pion mass 
could be attained.

With the vacuum polarization functions $\Pi_r(s)$ being ready in the whole $s$ range
and the relation $\Pi_{\rm EM}(s)=\sum_{r=\rho, \omega, \phi} \Pi_{r}(s)$,
we get the HVP contribution through Eq.~(\ref{amu4})
\begin{eqnarray}
a^{\rm HVP}_\mu= (641^{+65}_{-63})\times 10^{-10},\label{va}
\end{eqnarray}
to the muon anomalous magnetic moment, where the uncertainty comes from the variation
of the pertubative inputs by 10\%, and mainly from the $\rho$ channel.
The decomposition of the central value into the three pieces of resonance contributions gives $a^{{\rm HVP},\rho}_\mu  = 548\times 10^{-10}$, 
$a^{{\rm HVP},\omega}_\mu = 45\times 10^{-10}$, and $a^{{\rm HVP},\phi}_\mu = 49\times 10^{-10}$.
All the above results, consistent with those in the literature \cite{DellaMorte:2017dyu}, imply 
the success of our formalism: nonperturbative properties can be extracted from asymptotic QCD by 
solving an inverse problem. 
We remind that the result in Eq.~(\ref{va}) comes only 
from the considered $2\pi$, $3\pi$ and $KK$ channels. Adding the contributions from the other 
channels, such as $4\pi$ and charmonia, will increase our prediction for the HVP contribution.


\subsection{The Hybrid Approach}

A hybrid method has been proposed in \cite{Dominguez:2017yga}, which combines the data fitting and 
the LQCD input for the first derivative of the vacuum polarization function $\Pi^\prime_{\rm EM}(0)$. 
The final expression for the light-quark HVP contribution to the muon anomalous magnetic moment is 
written as
\begin{eqnarray}
a_\mu^{\rm HVP}=183.2\pm 2.1+5027\Pi^\prime_{\rm EM}(0)\; \mathrm{GeV}^2\; [10^{-10}],\label{Eq:amu}
\end{eqnarray}
where the first error largely stems from the data of the $e^+e^-$ annihilation
cross section. The first derivative in the second term is given by the sum 
$\Pi^\prime_{\rm EM}(0)=\sum_{r=\rho, \omega, \phi} \Pi^\prime_{r}(0)$ with each piece
\begin{eqnarray}
\Pi^\prime_r(0)&=&\int_{\lambda_r}^{\Lambda_r}\frac{\mathrm{Im}\Pi_r(s)}{s^2}ds+
\int_{\Lambda_r}^{\infty}\frac{\mathrm{Im}\Pi_r(s)}{s^2}ds,
\label{Eq:pi}\end{eqnarray}
where the determined parameters in Eq.~(\ref{Eq:fitres}) are taken for the first integral, and 
the perturbative input is inserted into the second integral.
Equation~(\ref{Eq:pi}) then yields the first derivatives at the origin
\begin{eqnarray}
\Pi^{\prime}_\rho(0)=0.0819,\quad
\Pi^\prime_\omega(0)=0.0063,\quad
\Pi^{\prime}_\phi(0)=0.0066,
\label{Eq:sep}
\end{eqnarray}
which are scheme-independent, though the on-shell scheme has been adopted. 
Substituting Eq.~(\ref{Eq:sep}) into Eq.~(\ref{Eq:amu}), we have 
\begin{eqnarray}
a_\mu^{\rm HVP}=(660\pm2^{+48}_{-48})\times 10^{-10}.
\end{eqnarray}
This value, turning out to be close to that in \cite{Davier:2019can},  
further supports our formalism for evaluating the vacuum polarization.
The accuracy of a calculation in the hybrid approach can be improved by including higher derivatives
of the vacuum polarization function \cite{Dominguez:2017yga}, which are not yet available in LQCD, 
but can be derived using our formalism.

At last, we present an alternative expression for the vacuum polarization function, which may be considered for a hybrid approach. Starting with Eq.~(\ref{dis0}) and following the idea 
of \cite{Dominguez:2017yga,Groote:2003kg,Dominguez:2017omw}, we write
\begin{eqnarray}
\Pi_{\rm EM}(-s)=\frac{1}{2\pi i}\oint_{|s'|=\Lambda} ds'\frac{\Pi_{\rm EM}(s')}{s'}
+\frac{1}{\pi}\int_{s_{\rm thr}}^\Lambda ds'\frac{{\rm Im}\Pi_{\rm EM}(s')}{s'+s}
-\frac{s}{\pi}\int_\Lambda^\infty ds'\frac{{\rm Im}\Pi_{\rm EM}(s')}{s'(s'+s)},
\end{eqnarray}
with the threshold $s_{\rm thr}$.
The first and third terms can be computed in perturbation theory for
a large enough scale $\Lambda$, and the second term, receiving the low
mass contribution, can take the data input. 

\subsection{Analyticity Constraint}

\begin{figure}
\subfigure[]{\includegraphics[scale=0.47]{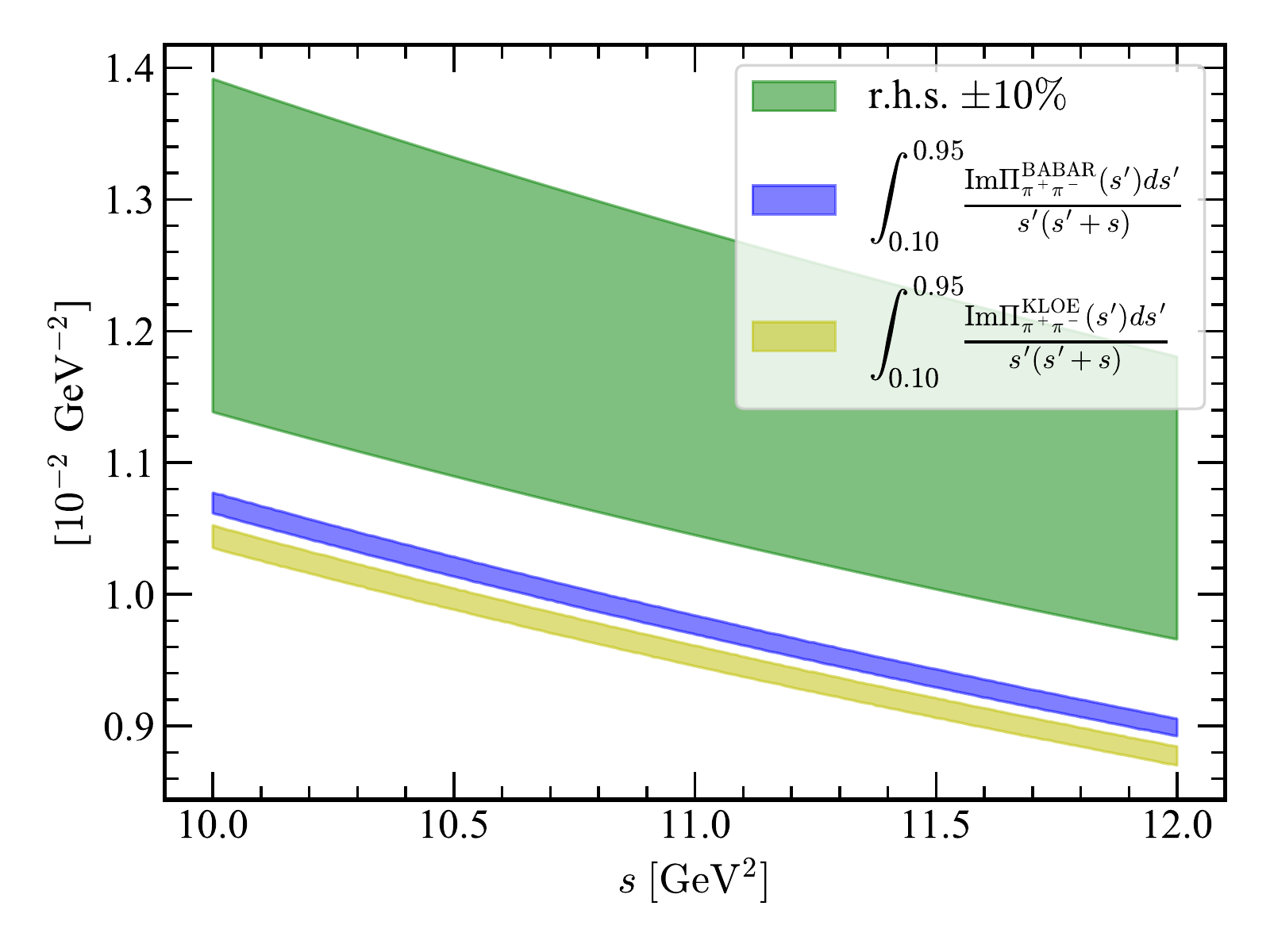}}
\subfigure[]{\includegraphics[scale=0.25]{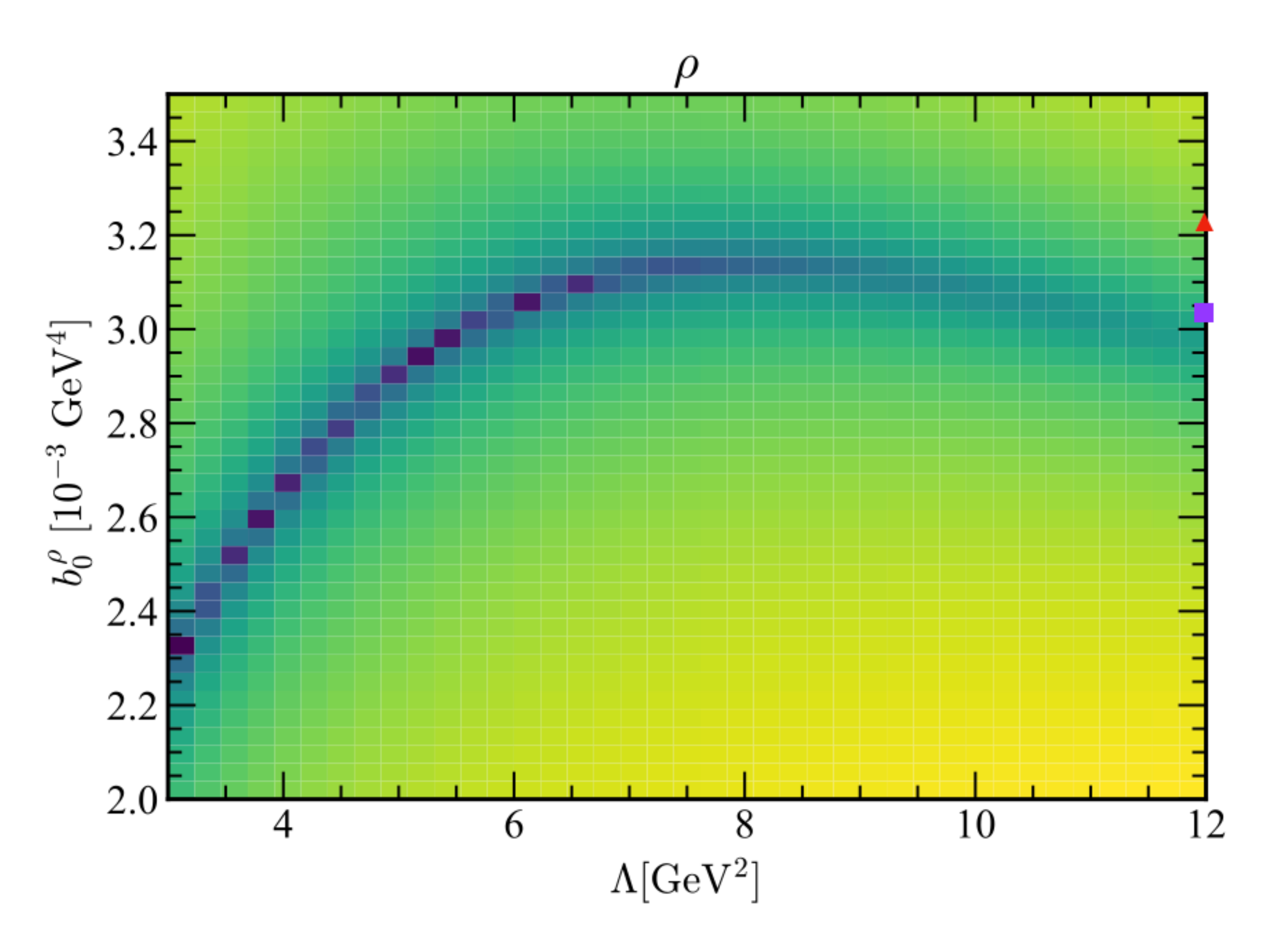}}
\caption{\label{fig6}
(a) Low-energy experimental data confronted with the analyticity constraint. The blue and yellow bands show $1\sigma$ errors estimated from the correlated uncertainty in the BABAR and KLOE data, respectively, while the green band represents the r.h.s. of Eq.~(\ref{Eq:new1}) with the 10\% uncertainty from the perturbative input. 
(b) Minimum distribution of RSS on the $\Lambda-b_0^\rho$ plane. The red triangle and the purple square denote the $\Lambda$ and $b_0^\rho$ values that best fit the KLOE and BaBar data, respectively.}
\end{figure}

As stated in the Introduction, simply inputting data into a dispersive approach does not automatically guarantee exact 
realization of the analyticity. Note that the perturbative $\mathrm{Im}\Pi(s)$ has been employed to evaluate the $R$-ratio, $R_{\rm QCD}$, for $\sqrt{s}>1.8$ GeV in \cite{Davier:2019can}. To satisfy the analyticity constraint, the dispersive integral of a dataset at low energy must reproduce the real part of the vacuum polarization function $\Pi(s)$ at large $s$. However, this self-consistency has never been examined seriously in the literature. 
Here we briefly demonstrate how to discriminate the BABAR \cite{Aubert:2009ad, Lees:2012cj} and KLOE \cite{Anastasi:2017eio} data for $e^+e^-\to \pi^+\pi^-$ by imposing the analyticity constraint, although a rigorous discrimination requires more precise perturbative inputs. For the latter, the results in 2008 \cite{Ambrosino:2008aa}, 2010 \cite{Ambrosino:2010bv} and 2012 \cite{Babusci:2012rp} have been combined. 
The dispersion relation in Eq.~(\ref{disy}) for $r=\rho$ is rewritten as
\begin{eqnarray}
\int_{0.10}^{0.95}\frac{\mathrm{Im}\Pi_{\pi^+\pi^-}^{\mathrm{BABAR, KLOE}}(s^\prime)ds^\prime}{s^\prime(s^\prime+s)}&=&\Omega_\rho(s) +\pi\frac{\Pi_\rho(0)}{s} - \int^{\Lambda_\rho}_{0.95}\frac{\mathrm{Im}\Pi_\rho(s^\prime)ds^\prime}{s^\prime(s^\prime+s)},\qquad\label{Eq:new1}
\end{eqnarray}
where the range of $0.10~\mathrm{GeV}^2<s^\prime<0.95~\mathrm{GeV}^2$ is the common domain of the BABAR and KLOE data. On the left-hand side of Eq.~(\ref{Eq:new1}), the BABAR and KLOE data for $\pi^+\pi^-$ are converted into $\mathrm{Im}\Pi(s^\prime)$, and the integrals for $10~\mathrm{GeV}^2<s<12~\mathrm{GeV}^2$, approximated by discretized sums, are presented in Fig.~\ref{fig6}(a). These integrals represent the contributions from the BABAR and KLOE data to the right-hand side of Eq.~(\ref{Eq:new1}). The discrepancy between the BABAR and KLOE bands implies that these two datasets cannot respect the analyticity constraint simultaneously. 
The $2.5\%$ difference between the central values of the two dispersive integrals persists to the higher $s$ region. The same amount of difference has been observed between the contributons to the muon $g-2$ from the BABAR and KLOE data in the phenomenological approaches \cite{Davier:2019can, Keshavarzi:2019abf}.
We have also computed the dispersive integral for the
SND data, which, if included into Fig.~\ref{fig6}(a), is located between and overlaps with the BABAR and KLOE bands.

Next we adopt the perturbative input for $\Omega_\rho$ in Eq.~(\ref{disy}), and the solution which respects the analyticity, ie., the parameters determined in Eq.~(\ref{Eq:fitres}) for $\Pi_\rho(0), \mathrm{Im}\Pi_\rho$ and $\Lambda_\rho$ to get the right-hand side of Eq.~(\ref{Eq:new1}). The right-hand side estimated at leading order with the 10\% uncertainty gives the wide band above the BABAR one, indicating that the BABAR data, whose dispersive integral is closer to the solution, are more favored over the KLOE and SND data by the analyticity requirement. 
To discriminate the BABAR and KLOE data, the evaluation of the right-hand side of Eq.~(\ref{Eq:new1}) should be more precise than 2.5\%. For $\mathrm{Im}\Pi(s)$ (or equivalently the $R$-ratio) in the definition of $\Omega_\rho(s)$, the calculation has been performed up to $\alpha_s^4$ \cite{Baikov:2008jh}, thus being precise enough: the precision of $R_{\rm QCD}$ has reached about 0.5\% according to \cite{Davier:2019can} for the range $15~\mathrm{GeV}^2<s^\prime<250~\mathrm{GeV}^2$, where the inputs to our analysis are selected. In principle, the real part of $\Pi(s)$ should be computed up to the same order for consistency, and a precision of 0.5\% is expected. Then $\Omega_\rho(s)$ will be determined precisely enough, with which we can also update the second and third terms on the right-hand side of Eq.~(\ref{Eq:new1}) to the same precision by solving the Fredholm equation. We conclude that it is possible to discriminate the BABAR and KLOE data with the $2.5\%$ difference by higher order calculations for $\Pi(s)$ in the large $s$ region.

As emphasized before, the analyticity constraint imposes a correlation among
the parameters involved in Eq.~(\ref{para}). For $\Lambda_\rho$ and $b_0^\rho$,
their correlation is described by the minimum distribution of RSS on the $\Lambda_\rho$-$b_0^\rho$ plane
in Fig.~\ref{fig6}(b). This minimum distribution is equivalent to that in Fig.~\ref{fig2}(a),
but projected on to the $\Lambda_\rho$-$b_0^\rho$ plane. Ignoring the correlation and simply 
fitting ${\rm Im}\Pi_\rho(s)$ to the data, as done in the conventional dispersive approach, we
find $\Lambda_\rho$ and $b_0^\rho$ marked by the square and triangle in Fig.~\ref{fig6}(b)
for the BABAR and KLOE data, respectively. The distance
between a mark and the RSS minimum distribution reflects the deviation of
the corresponding dataset from the analyticity constraint. It is obvious that
the BABAR dataset, being nearer to the minimum distribution than the KLOE one, 
respects more the analyticity constraint, an observation consistent with the indication of
Fig.~\ref{fig6}(a). To realize our proposal by means of the conventional dispersive approach,
one can assign a weight with each dataset in the fit according to its 
distance to the minimum distribution. Certainly, the analysis will be lengthier due to
the more complicated model for the resonance spectra in \cite{Davier:2019can}:
one has to derive the minimum distribution, determine the best-fit points for the adopted datasets,
and assign weights according to the distances between them in the multi-dimensional space formed by 
the involved parameters. 
If it turns out that the KLOE data are not favored by the analyticity requirement with sufficiently 
precise perturbative inputs, the removal of the KLOE dataset from the fit will enhance the $\pi\pi$ contribution 
to $a_\mu^{\rm HVP}$ from $507.9\times 10^{-10}$ up to $510.6\times 10^{-10}$ \cite{Davier:2019can}. That is, 
the central value of $a_\mu^{\rm SM}$ could be increased by $\sim 3 \times 10^{-10}$. 
Given that the theoretical precision of $a_\mu^{\rm SM}$ is unchanged, the anomaly could be reduced from $3.3\sigma$ to $3.0\sigma$. This reduction elaborates the potential impact of our work.


\section{CONCLUSION}

In this paper we have extended a new formalism for extracting nonperturbative observables  
to the study of the HVP contribution $a^{\rm HVP}_\mu$ to the muon anomalous magnetic moment $g-2$.
The dispersion relation for the vacuum polarization function $\Pi(q^2)$ was turned into 
an inverse problem, via which $\Pi(q^2)$ at low $q^2$ was solved with the perturbative 
input of $\Pi(q^2)$ at high $q^2$. Though multiple solutions exist, the best ones can be 
selected, which accommodate the data of the $e^+e^-$ annihilation cross section. Because 
the involved parameters are correlated under the analyticity requirement of the vacuum 
polarization, and not completely free, the satisfactory agreement of our solutions 
with the data is nontrivial.  It has been shown that our prediction for $\Pi(q^2)$, 
including its first derivative $\Pi^\prime(0)$, is close to those from LQCD, and contributes
$a^{\rm HVP}_\mu= (641^{+65}_{-63})\times 10^{-10}$ to the muon $g-2$ from the 
$\rho$, $\omega$, and $\phi$ resonances in consistency with the observations from the other 
phenomenological, LQCD and hybrid approaches. The slight deviation 
of our result for the $\rho$ resonance spectrum from the SND data could be resolved by
considering subleading contributions to the perturbative input. This subject will be
investigated systematically in a forthcoming publication, and the corresponding theoretical
uncertainty is expected to be reduced. Other sources of uncertainties need to be examined, 
such as the one from different parametrizations for the resonance spectra.

The purpose of this work is not to fit the $e^+e^-$ annihilation 
data exactly, but to demonstrate how our formalism is implemented, and that reasonable
results can be produced even with a simple setup like the leading order perturbative input, 
the naive parametrizations in Eq.~(\ref{para}), and the fit only to the SND data. 
We stress that imposing the analyticity constraint to
the conventional phenomenological approach, which solely relies on data fitting, 
forms a more self-consistent framework for determining $a^{\rm HVP}_\mu$ in the Standard Model
with higher precision. We have explained how to discriminate the BABAR and KLOE 
data for $e^+e^-\to \pi^+\pi^-$ via the analyticity constraint as an example,
and proposed to assign weights with fitted datasets according to their deviation 
from the solutions of the inverse problem.
The success achieved in this paper also stimulates further applications of our formalism to 
the hadronic contributions to the muon $g-2$ from heavy quarks and from the 
light-by-light scattering \cite{deRafael:1993za,Hong:2009zw,Cappiello:2010uy,
Hagelstein:2017obr}, for which a lack of experimental information persists, and a theoretical 
estimation is crucial.

\vskip 1.0cm

{\bf Acknowledgement}\\
The authors would like to thank Fanrong~Xu and Fu-sheng Yu for useful comments.
This work was supported in part by MOST of R.O.C. under Grant No.
MOST-107-2119-M-001-035-MY3.


\end{document}